\documentclass{iaus}
\usepackage{graphicx}

\title[ULXs in the Cartwheel] 
{Variability of ultraluminous X-ray sources in the Cartwheel Ring}

\author[A. Wolter, G. Trinchieri and M. Colpi]   
{Anna Wolter$^1$, 
 Ginevra Trinchieri,$^1$ \break \and Monica Colpi$^2$}

\affiliation{$^1$INAF, Osservatorio Astronomico di Brera, via Brera 28, 20121 Milano, Italy \break email: anna.wolter@brera.inaf.it, ginevra.trinchieri@brera.inaf.it\\[\affilskip]
$^2$Dipartimento di Fisica G. Occhialini, 
Universit\`a degli Studi di Milano Bicocca, \break Piazza della Scienza 3,
20126 Milano, Italy \break email: Monica.Colpi@mib.infn.it}

\pubyear{2006}
\volume{238}  
\pagerange{001--999}
\date{??? and in revised form ???}
\jname{Black Holes: from Stars to Galaxies -- across the Range of Masses}
\editors{V. Karas \& G. Matt, eds.}

\def\ergs{$\rm {erg\,s}^{-1}$}
\def\sun{$_{\odot}$}

\begin{document}

\maketitle

\begin{abstract}
The Cartwheel is one of the most outstanding examples of a dynamically
perturbed galaxy where star formation is occurring inside the
ring--like structure.  In previous studies with {\it Chandra}, we detected
16 Ultra Luminous X-ray sources lying along the southern portion of
the ring. Their Luminosity Function is consistent with them being in the high
luminosity tail of the High Mass X-ray Binaries distribution, but with one 
exception: source N.10.  
This source, detected with {\it Chandra} at $L_X = 1 \times
10^{41}$\ergs, is among the brightest non--nuclear 
sources ever seen in external galaxies.  Recently, we have observed
the Cartwheel with {\it XMM-Newton} in two epochs, six months apart.
After having been at its brightest for at least 4 years, the source has 
dimmed by at least a factor of two between the two
observations. This fact implies that the source is compact
in nature. Given its extreme isotropic luminosity, there is the possibility 
that the source hosts an accreting 
intermediate--mass black hole.  Other sources in the ring vary in flux
between the different datasets.  We discuss our findings in the
context of ULX models.  

\end{abstract}

\firstsection 
\section{Introduction}

Very luminous off-nuclear X-ray sources in nearby galaxies
are known since the {\it Einstein}
satellite times.
They have been named Ultra-Luminous X-ray sources (ULXs) because
their isotropic X-ray luminosity is significantly higher 
than the Eddington 
limit for a solar mass black hole ($L_X \sim  1.4 \times 10^{38}$ \ergs\,).
The name itself is only a phenomenological description of their
$L_X$, since their nature is not clear yet.
They do not appear to have an equivalent among Galactic sources, and
this may be related to the low Star-Formation rate of the Milky Way
since ULXs are mostly found associated with Star-Forming regions.
Explanations of their nature involve beamed emission from an accreting stellar 
mass compact object, or super-Eddington emission, or isotropic 
accretion onto an intermediate--mass black hole.

An extraordinary example of ULX is the source N.10 detected
in the narrow, gas-rich star--forming ring of the Cartwheel galaxy
with isotropic luminosity of $L_{0.5-10 \rm{keV}} \sim  
1.3 \times 10^{41}$ \ergs\, (Wolter et al. 1999; 
Wolter \& Trinchieri 2004 - hereafter WT04; Gao et al. 2003). 
This is the brightest of a number of very bright individual sources that 
also appear to reside in the ring, with 
isotropic luminosities in excess 
of $L_{0.5-10{\rm keV}} = 3 \times  10^{39}$ \ergs\, (WT04).

Source N.10 is one of the brightest ULXs known, with observed $L_X$
comparable with the peak $L_X$ of the 
best studied example of the class,
source X-1 in M82 (Ptak \& Griffiths 1999; Matsumoto et al. 2001; 
Kaaret et al. 2001; Strohmayer \& Mushotzky 2003; Dewangan et al. 2006).
The spectral and temporal variability of X-1 (typically bursts
of about a month duration) have led to the estimate of
200M\sun\, for the accreting object responsible for the X-ray emission
(Dewangan et al. 2006).

The lack of detailed information on variability for source N.10 
in the Cartwheel has instead prevented up to now 
the exclusion of an extended nature. We  present here 
new {\it XMM--Newton} observations which 
have now confirmed its compact nature.

\section{{\it XMM--Newton} observations}\label{sec:xmm}

\begin{figure}
\begin{center}
\resizebox{8cm}{!}{\includegraphics{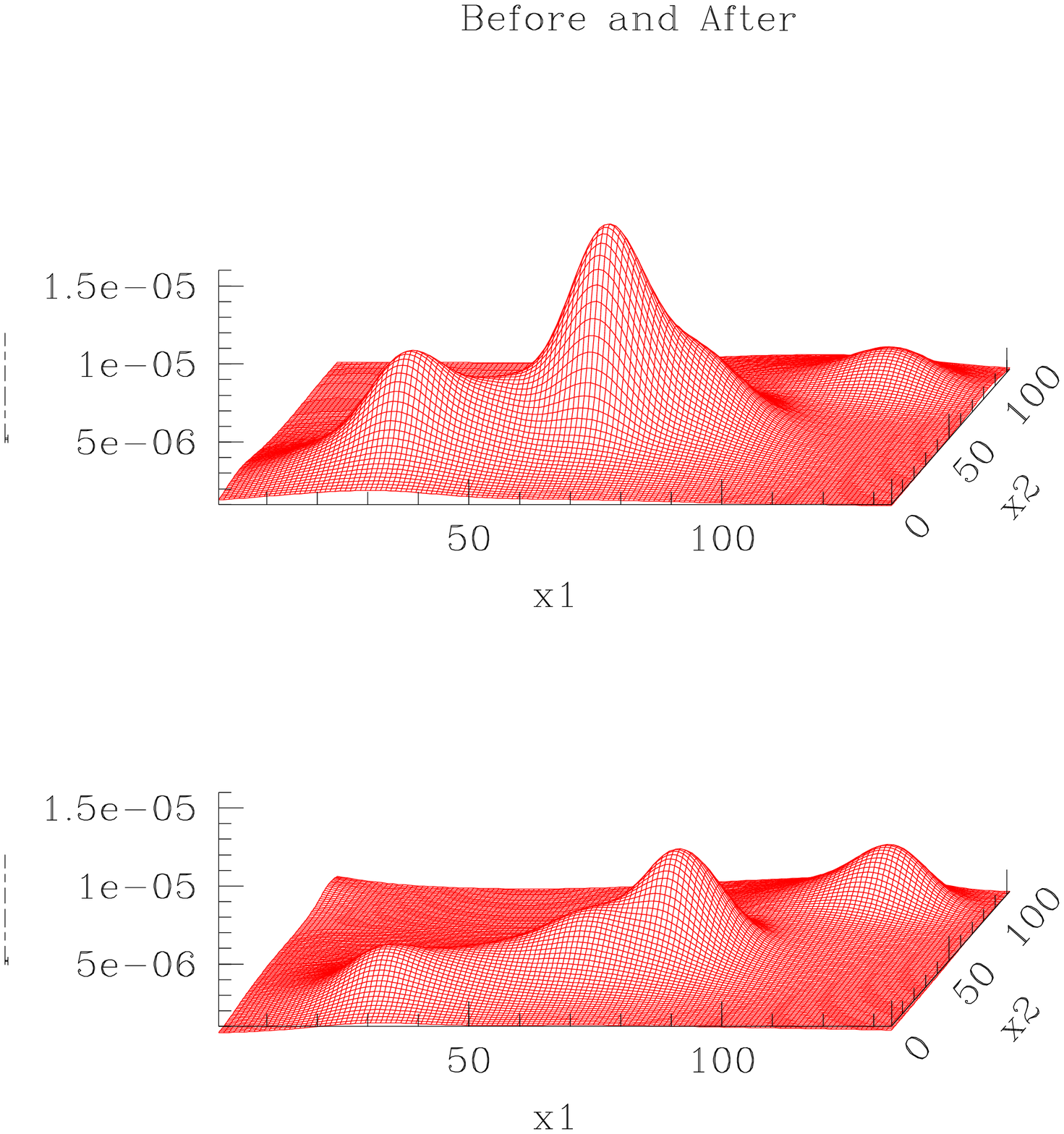} }
\end{center}
  \caption{A 3-d representation of the two {\it XMM-Newton} pn
observations [101] (top) and [201] (bottom) with a six months 
time separation. The smoothed images have been normalized to the
respective exposure time. It is evident that the peak located
at the position of source N.10 is very less prominent in the 
second epoch.
}\label{fig:surf}
\end{figure}

We have observed the source with {\it XMM-Newton} in two epochs
(December 2004 and May 2005).  A detailed description of the results
obtained for source N.10 is presented
in Wolter, Trinchieri \& Colpi (2006).
A 3-d representation of the two smoothed datasets is plotted in 
Figure~\ref{fig:surf}. 

\begin{figure}[htb]
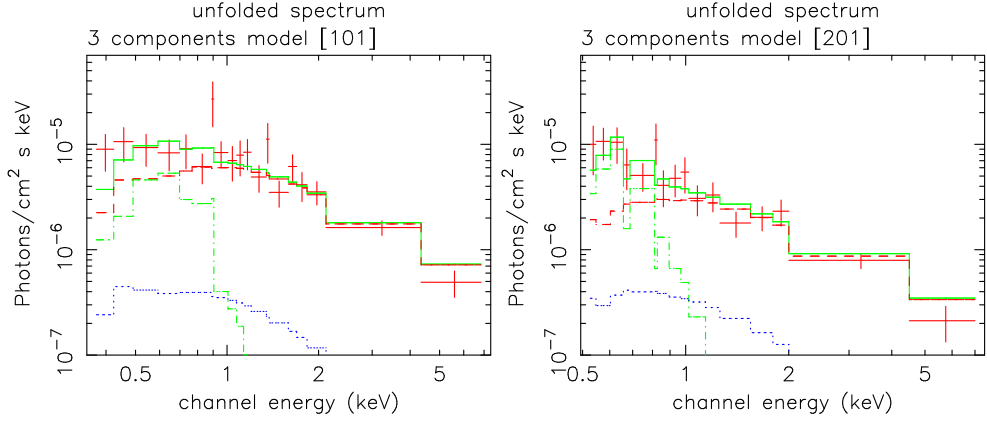

\hbox{
\includegraphics[width=5.5cm,angle=270]{fig2a.eps} 
\includegraphics[width=5.5cm,angle=270]{fig2b.eps}
}
\caption{Unfolded spectrum,
from pn data only for clarity, showing the three components (green
dot-dashed line = gas; blue short dashed line = unresolved binaries; 
red long dashed line = ULX) at the best
fit values. The difference between {\it left} [101] and {\it right} [201]
is only the normalization of the `ULX' component [besides the
binning scheme due to the different statistics at the two epochs]. }
\label{fig:usf}
\end{figure}

The detection of variability is important because it is our strongest 
proof that we are detecting a single 
source, namely an accreting binary, and not a collection of less
luminous unresolved X-ray sources. It is also relevant because
it can provide
essential information about the details of the accretion process 
and, most important, about the masses involved in the
process (donor and accretor), giving the possibility to infer the
presence of an intermediate-mass black hole.

The spectrum in the two different {\it XMM-Newton} observation (the first,
[101] of 24 ksec, and the second [201], of 42 ksec exposure) is
shown in Figure~\ref{fig:usf}.
The {\it XMM-Newton} PSF is such that the extraction radius (even if we
used a smaller than customary value of 10$^{\prime\prime}$) contains 
part of the ring emission, which
is described by the sum of a diffuse gas component (thermal spectrum)
and the contribution from the unresolved point-like sources (X-ray binaries; 
power law spectrum. See WT04 for details).
Even if the statistics is low, therefore, a multicomponent spectrum
has been used to fit the emission: we have added a power law describing
the ULX power law emission (as seen in {\it Chandra}) to the two ring
components. The spectra of the two observations 6 months apart 
clearly show that the thermal component (gas in the ring) is constant
in time and that the factor of 2 variability of the ULX is
quite evident.

\begin{figure}[htb]
\hbox{
\includegraphics[width=6.5cm]{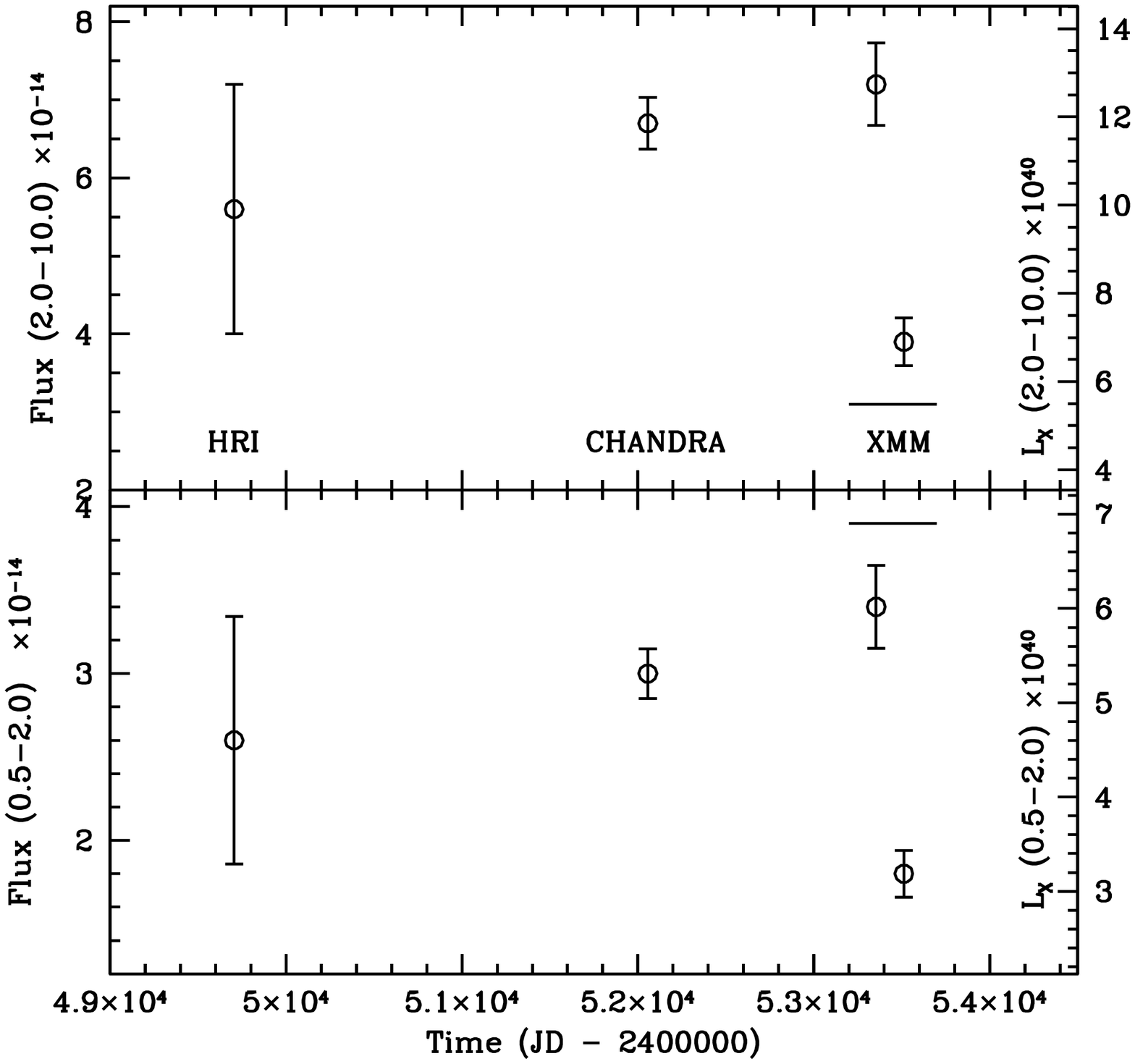} 
\includegraphics[width=6.5cm]{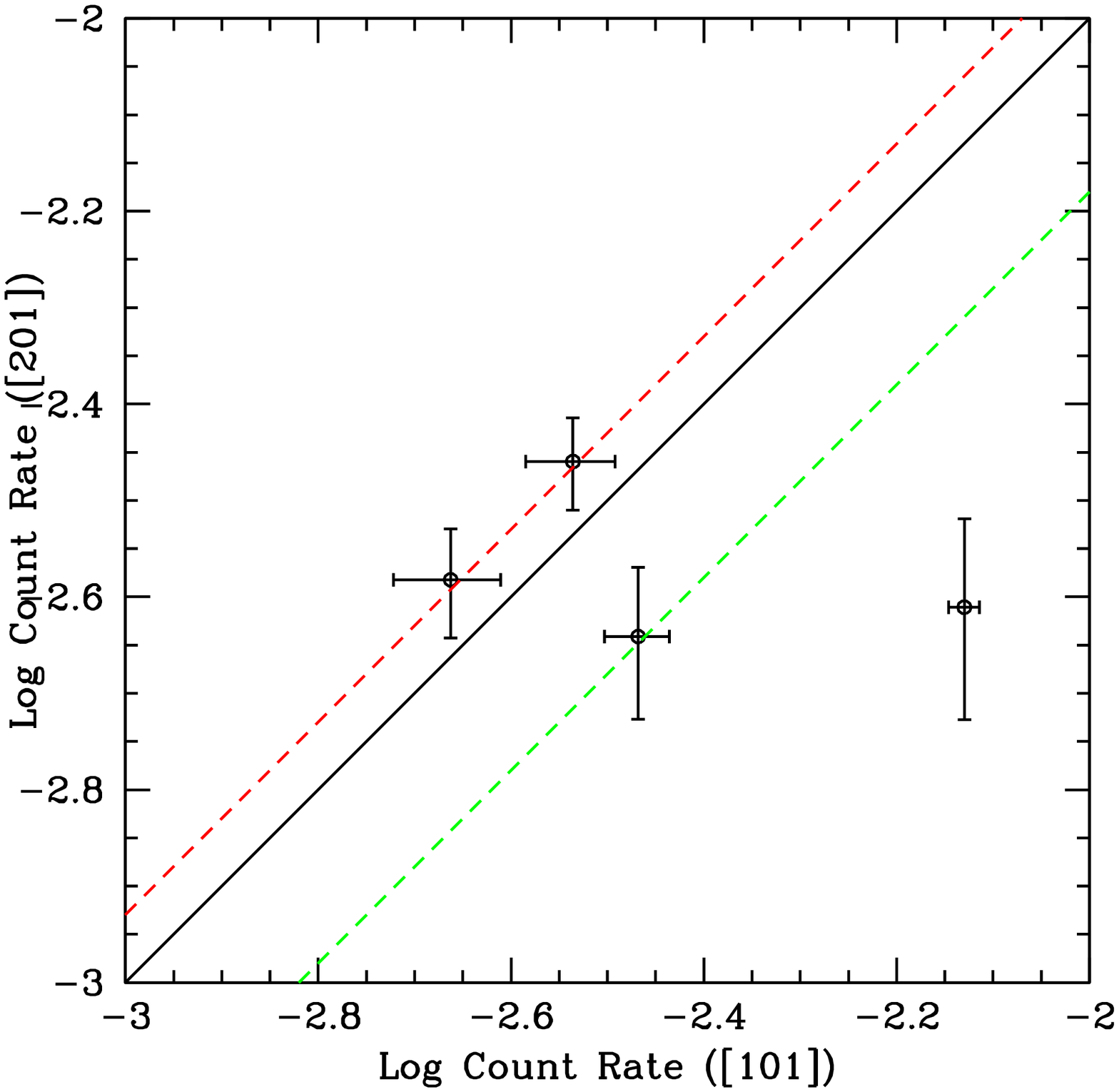} 
\hfill
}
 \caption{{\it Left:} Long term light curve in the soft (lower panel) and hard
 (upper panel) energy
 band, over an interval of about 10 years. The two {\it XMM--Newton} points 
that define the variation are
 not subject to cross-calibration uncertainties. All fluxes are computed 
 with the same spectral shape,
 i.e. power law with $\Gamma = 1.6$ and N$_{\rm H} = 3.6 \times 10^{21}$ 
 cm$^{-2}$; see fit to {\it Chandra} data in WT04. Right axis reports 
luminosities computed assuming the Cartwheel distance. From Wolter et al. 2006.
{\it Right:} Flux comparison of the brightest sources in the first
and second epoch.}

\label{fig:lc}
\end{figure}
The long term behavior of N.10 shows a nearly constant luminosity
for about 10 years (see Figure 3 -- {\it Left}), and
then a rapid dimming of at least a factor of 2 in 6 months.
This is quite different from the variability pattern observed in
the best studied bright ULX, namely X--1 in the starburst galaxy M82
(Dewangan et al. 2006).

We compared count rates for the brightest 
source neighboring N.10. The comparison is shown in Figure 3 ({\it Right}).
We have determined that the source next to it,
which corresponds to {\it Chandra} sources N.13 \& N.14, has not varied between
the two observations (the count rate in the second observation is at most 15\%
higher than in the first one). If we assume that this is constant, then
the source to the NW, corresponding to N.16 \& N.17, is also constant (the
same 15\% increase in the count rate), but the SE source (N.7 \& N.9) 
has faded to
about half its strength in the second observation.  

\section{Conclusions}\label{sec:concl}
The whole ring of the
Cartwheel consists of bubbles and condensations (Struck et al. 1996) and
the neighborhood of N.10 is no exception; 
the association with an environment of massive and young
stars is almost certain.
The most appealing interpretation for the emission of source N.10
is that it is powered by an intermediate--mass black hole. 
We cannot rule out the presence of Super-Eddington accretion.
Other sources have been observed to vary in the same time frame of
N.10. 
Further studies on variability will help us in distinguishing between
different accretion modes. It is evident that the Cartwheel ring
is an excited (and exciting!) environment, in which star formation
is violently at work. The low metallicity measured for the optical 
gas might be a key ingredient for the formation of sources capable of
emitting very high X-ray luminosity, whatever the mechanism is.

\begin{acknowledgments}
We  acknowledge partial financial support from the Italian
Space Agency (ASI) under contract ASI-INAF I/023/05/0.
This research has made use of SAOImage DS9, developed by the Smithsonian
Astrophysical Observatory.
This work is based on observations obtained with {\it XMM-Newton}, an ESA
science mission with instruments and contributions directly funded by
ESA Member States and the USA (NASA).
\end{acknowledgments}

\end{document}